\documentclass[sn-mathphys,Numbered]{sn-jnl}% Math and Physical Sciences Reference Style
\usepackage{graphicx}%
\usepackage{multirow}%
\usepackage{cite}
\usepackage[table]{xcolor}
\usepackage{arevmath}
\usepackage{amsmath,amssymb,amsfonts}
\usepackage{amsthm}%
\usepackage{mathrsfs}%
\usepackage[title]{appendix}%
\usepackage{xcolor}%
\usepackage{textcomp}%
\usepackage{manyfoot}%
\usepackage{booktabs}%

\begin{document}

\title[Fuzzy Gene Selection and Cancer
Classification Based on Deep Learning
Model]{Fuzzy Gene Selection and Cancer
Classification Based on Deep Learning
Model}

\author*[1,3]{\fnm{ MAHMOOD} \sur{KHALSAN}}\email{mahmood.khalsan@northampton.ac.uk}

\author[1]{\fnm{MU } \sur{MU,(Member, IEEE)}}\email{mu.mu@northampton.ac.uk}
\equalcont{These authors contributed equally to this work.}

\author[3]{\fnm{EMAN SALIH } \sur{AL-SHAMERY}}\email{emanalshamery@itnet.uobabylon.edu.iq}
\author[2]{\fnm{LEE  } \sur{MACHADO}}\email{lee.machado@northampton.ac.uk}
\author[1]{\fnm{SURAJ  }\sur{AJIT}}\email{suraj.ajit@northampton.ac.uk}
\author[1]{\fnm{MICHAEL  }\sur{ OPOKU AGYEMAN , (Senior Member, IEEE)}}\email{michael.opokuagyeman@northampton.ac.uk}

\equalcont{These authors contributed equally to this work.}

\affil*[1]{\orgdiv{Advanced Technology Research Group}, \orgname{Faculty of Arts, Science and Technology}, \orgaddress{\street{The University of Northampton}, \country{UK}}}

\affil[2]{\orgdiv{Centre for Physical Activity and Life Science}, \orgname{Faculty of Arts, Science and Technology}, \orgaddress{\street{The University of Northampton}, \state{UK}, \country{Country}}}

\affil[3]{\orgdiv{Computer Science Department}, \orgname{University of Babylon},   \country{Iraq}}
\abstract{Machine learning (ML) approaches have been used to develop highly accurate and efficient
applications in many fields including bio-medical science. However, even with advanced ML techniques,
cancer classification using gene expression data is still complicated because of the high dimensionality of the
datasets employed. We developed a new fuzzy gene selection technique (FGS) to identify informative genes
to facilitate cancer classification and reduce the dimensionality of the available gene expression data. Three
feature selection methods (Mutual Information, F-ClassIf, and Chi-squared) were evaluated and employed
to obtain the score and rank for each gene. Then, using Fuzzification and Defuzzification methods to obtain
the best single score for each gene, which aids in the identification of significant genes. Our study applied
the fuzzy measures to six gene expression datasets including four Microarray and two RNA-seq datasets
for evaluating the proposed algorithm. With our FGS-enhanced method, the cancer classification model
achieved 96.5\%,96.2\%,96\%, and 95.9\% for accuracy, precision, recall, and f1-score respectively, which is
significantly higher than 69.2\% accuracy, 57.8\% precision, 66\% recall, and 58.2\% f1-score when standard
MLP method was used. In examining the six datasets that were used, the proposed model demonstrates its capacity to  classify cancer effectively.}

\maketitle
\section{Introduction}
Cancer is the second leading cause of death worldwide and represents  the abnormal growth of cells and their frequent metastatic spread throughout the body [1]. Cancer cells frequently proliferate independently of growth signals. and neglect to respond to survival/death that instructs them to stop dividing or to die (i.e. by apoptosis). This phenomenon occurs due to inherited or environmental factors that cause DNA mutations or epigenetic modifications that deregulate normal cellular gene expression programs [2]. For example, DNA mutation is caused by harmful substances in the environment including chemicals in tobacco smoke and ultraviolet radiation from the sun. Some cancer genes are inherited (i.e. BRCA1/2) and have high penetrance due to their fundamental role in cellular regulation. Therefore, the analysis of deregulated gene expression programs in cancer cells may play an important role in the early detection and treatment of cancer. Consequently, identifying a specific set of genes (gene signatures) that aid classification may provide an earlier diagnosis of cancer and provide personalized treatment options [2]. The tools (Microarray and RNA-seq technologies) that have been developed for measuring the expression levels of genes in normal and cancer tissue have opened the door for investigators to build and test a new mathematical and statistical model for analyzing gene expression data. Those measurement tools calculate the expression levels of thousands of genes across hundreds/thousands of clinical samples.\\ 
Both (Microarray and RNA- seq technologies)  measure transcriptome-wide gene expressions and allow a comparison of cancerous and non-cancerous tissues. Microarray methods measure the intensities of colored fluorescent probes spotted on glass slides, which correspond to gene expression under different conditions. Whereas RNA-Seq methods measures read counts as a proxy for relative gene abundance [3]. RNA-seq methods have largely superseded microarrays as they produce less noise and are more accurate in calculating method gene expression abundance [4]. Researchers have developed a range of mathematical and statistical techniques to analyze gene expression data for various goals. This includes the identification of optimal gene signature pathways, enhanced cancer classification, cancer prediction, drug discovery, and improved personalized therapy. To achieve this, obstacles regarding the high dimensionality and complexity of the publicly available gene expression data remain. However, measurement tools for calculating gene expressions have improved continuously. Artificial intelligence (AI) is now a powerful tool for mitigating the time taken to analyze large cancer datasets. It has the potential to improve the accuracy of cancer classification and/or cancer prediction. AI is the broadest term used to classify machines that mimic human intelligence. AI includes machine learning (ML) techniques including Support Vector Machine (SVM), K-Nearest Neighbour (KNN), and Random Forest (RF) approaches. ML also includes deep learning (DL) approaches that use Convolutional Neural Networks (CNN), Long short-term memory (LSTM), and, MLP.\\

The present study provides significant contributions by attempting to address a number of shortcomings.\\
First, a new fuzzy gene selection technique has been developed to make the datasets on gene expression less dimensional.\\
Second, using a limited number of genes when using the FGS method prevents or at least reduces overfitting problems when classifier approaches are applied.\\
Third: Reducing the amount of time required for a classifier model's training stage is made possible by a minimal number of biomarker genes that are utilized as identifiers.\\
Fourth: The suggested paradigm enables early cancer detection and precise cancer classification.\\
Fifth: Choosing a few useful informative genes to be employed. \\

The rest of the work is organized as follows: section II explores recent studies analyzing gene expression data by using  ML. Section III explains theoretically the concepts of methods that have been used for developing the fuzzy gene selection methods and classifier approaches that have been employed. It also illustrated the repositories that have been used to download the datasets employed for training and testing the proposed model. While section IV  explains practically the techniques that have been employed for developing the proposed model (FGS and MLP). Section V discussed the results that have been obtained from the proposed model (FGS and MLP) and compared the other classifier approaches such as (i.e.SVM, KNN, and RF). Conclusions are provided at the end of the paper.

\section{Related Work}\label{sec2}
Sun et al. [5], suggested a new approach namely  a multimodel deep neural network (MDNN) that aims to improve the performance accuracy of breast cancer classification. The proposed algorithm was trained and tested on publicly available gene expression data that includes 24368 genes across 2509 breast cancer and 548 normal samples [6]. The new model was compared with three different machine learning methods (SVM, RF, and Logistic regression (LR)). Minimum Redundancy Maximum Relevance (mRMR) was also employed  as a feature selection Technique  to reduce  the number of features (genes) to improve the performance of classification accuracy. The accomplished accuracy was  82\%, 80\%, 79\% and 76\% for MDNN, SVM, RF, and LR respectively. However, recall values were low in all classifier algorithms (45\%, 36\%, 22\% and 18\% for MDNN, SVM, RF, and LR respectively) and precision was 95\% for all classifier approaches.

Although the suggested model's performance accuracy was good, further accuracy enhancement is necessary due to the cancer's sensitivity. Furthermore, the recall values were quite low, which had an impact on the performance of the provided method. Typically, research use several datasets for different types of cancer to validate the findings produced by applying their models, which have been evaluated in this work where just one dataset was used.

Jing Xu et al.  [7], proposed  a novel Deep Neural Forest (DFNForest) algorithm  to classify subtypes of three different cancer types (Glioblastoma multiforme (GBM)), Breast, and lung ). The system was tested by employing RNA-seq data available from TCGA. The researcher used two feature selection techniques (fisher ratio and neighborhood rough set) to reduce the dimensionality of the publicly available data, addressed overfitting issues, and selected the genes that significantly impacted the performance of the proposed model [8]. They achieved an accuracy of 93\% (breast), 88\% (lung), and 84\% (GBM).

Guillermo et al. [9], proposed CNN and transfer learning (TL) model for lung tumor prediction. (10535) samples
and the top 20k most expressed genes were downloaded from
TCGA for 33 different kinds of cancer but the proposed
model was tested only on the lung cancer dataset. The
system compared the new model against other classifier
methods(densely connected multi-layer feed-forward neural
network (MLNN) and SVM) to evaluate the suggested model.
The achieved accuracy was 68\%, 72\%, and 69\% for CNN,
MLNN and SVM respectively. The proposed model showed
that low accuracy was accomplished, and it was tested only on one
type of cancer(lung) that may not achieve the same score of
accuracy for other types of cancer. The proposed model
was not achieved better accuracy than compared classifier
methods that the investigator described in this study (MLNN)
was achieved better accuracy as illustrated previously. Other
evaluation measurements from this research were identified
in Table\ref{table2}.\\
 \begin{table}[htpb]
\caption{Comparing the performance of CNN against MLNN and SVM}
\label{table2}
\centering
\setlength{\tabcolsep}{5pt}
\begin{tabular}{lllll}
\hline
Methods& 
AUC &
Sensitivity&
Specificity &
Accuracy\\
\hline
CNN &
  \begin{tabular}[c]{@{}l@{}}73\%\end{tabular} &
  \begin{tabular}[c]{@{}l@{}}67\%\\ \end{tabular} &
 68\% & 68\%\\
MLNN &
  \begin{tabular}[c]{@{}l@{}}70\%\end{tabular} &
  \begin{tabular}[c]{@{}l@{}}61\%\\ \end{tabular} &
 73\% & 72\% \\ 
 SVM &
  \begin{tabular}[c]{@{}l@{}}70\%\end{tabular} &
  \begin{tabular}[c]{@{}l@{}}64\%\\ \end{tabular} &
 69\% & 69\% \\
\end{tabular}
\end{table}

Yeganeh et al. [10], multiple machine learning methods with multiple gene expression datasets of ovarian cancer employed for ovarian cancer prediction. Seven GEO datasets(GSE12172, GSE14407, GSE9899, GSE37648, GSE18521, GSE38666, and GSE10971) were obtained for training and testing the machine learning approaches. The system used a 26-gene set panel for training different classifier methods. The highest accomplished accuracy value was 0.89 when a Random Forest pipeline was applied. Low accuracy achieved and imbalanced datasets used were recorded as drawbacks in this work.

It concluded from this section that previous work requires developing a new model for improving cancer classification and selecting a small number of significant genes that would be used as identifiers for cancer classification. More studies were discussed in our previous published work freely available [30].
\subsection{Publicly available datasets}
Below  are common data repositories that provided gene expression data from  normal and cancer-derived tissues used to train and test models for classification or prediction purposes. 
Those repositories are further described as follows.
\subsubsection{Gene Expression Omnibus (GEO)} 
GEO [11] is a public functional genomics data repository supporting MIAME-compliant data submissions. The repositories support RNA-seq and Microarray data but GEO mostly provides Microarray data. The total number of samples that are provided by GEO is 3635328 for different diseases. GEO is freely available to download experiments and curated gene expression profiles by users or researchers.
\subsubsection{The Cancer Genome Atlas (TCGA)}
TCGA [12] is a landmark cancer genomics program that is distinguished in providing 84,031 samples from 33 different cancer types. The datasets that are available on TCGA are measured by the RNA-seq and Microarray methods for measuring expressed levels of gene activity for healthy and unhealthy tissues.
\subsection{Feature selection}
Feature Selection (FS) is a statistical method that aims to select an optimal feature of a large number of original features for given a dataset [13]. The goal is to choose the best subset of features with k features. FS approaches have valuable benefits in reducing the training time, reducing the complexity of the model, and are easy to interpret. Additionally, there are faster responses with unseen data and powerful generalization that enhances the performance of the model and avoids (or at least reduces) overfitting issues [14]. This work has used three feature selection methods to identify the optimal subset of genes that were employed later as identifiers for training classifier methods. Those feature selection methods are explained below.  
\subsubsection{Mutual Information }
Mutual information (MI) can be defined by how it gauges the amount of information shared by two random variables. In the context of gene selection, employs this definition to select a subset of important genes with respect to the output vector [14]. It has two major benefits: it can be used as a solution with different types of machine learning models, and it is a faster solution for selecting features. Mathematically it can be defined as follows. X represents the random variables(genes) and Y is the target (cancer types).\\
\begin{equation}
	I(X,Y) =\sum\sum p(X,Y)log \frac{p(x,y)}{p(x)p(y)}
	\label{Eq:I(X,Y)}
\end{equation}
\begin{equation}
	=H(Y)-H(Y/X)
	\label{s}
\end{equation}
Where H(Y|X) is the conditional entropy of Y in the case of X is known.
\subsubsection{ F-ClassIF}
\begin{figure}[htpb]
    \centering
\includegraphics[width=0.5\textwidth]{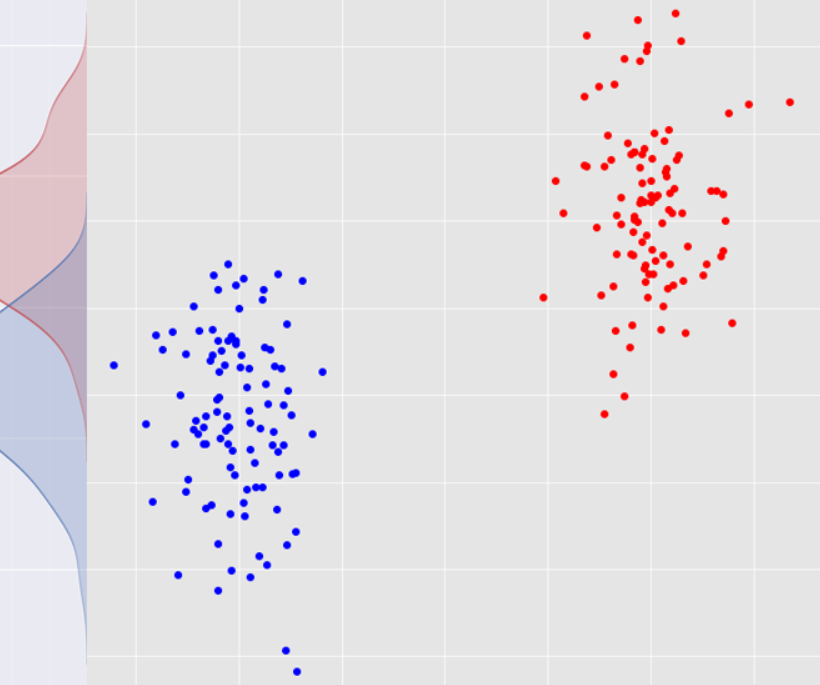}
\caption{Illustration example of  distributed Features to show up F-classif work}
    \label{fig1}
\end{figure}

F-class calculates the ratio between different values.
In other words, it calculates the variation between features/labels within the samples. This method is called the ANOVA f-test [15]. F-test results in a score that represents how far the feature is from other features. For example, calculate a score for each feature between two classes and use this score for selecting the important features. As shown in
In Figure\ref{fig1}, the red color presents class 1 and the blue color
introduces class 2 and two features on the x and y axes. The x feature is a better separator than y because if we project
data on the x-axis, two completely separated classes were
obtained but when project data onto y, two classes overlap
in the middle of the axis. Based on that the features which
were got higher scores will be chosen as the best features for
a given dataset.
\subsubsection{ Chi-squared}
 The chi-squared statistic is used to assess the independence of two occurrences. To begin, compute the chi-squared between each gene and the class. As a result, select the number of features based on the highest chi-squared scores. The chi-squared formula is presented below [16]: 
\begin{equation} \mathrm{X}_{\mathrm{c}}^{2}=\Sigma(\mathrm{O}_{\mathrm{i}^{-}}\mathrm{E}_{\mathrm{i}})^{2}/\mathrm{E}_{\mathrm{i}}
\end{equation}
Where:
C = degrees of freedom,
O = observed value(s), and
E = expected value(s)
\subsection{Fuzzy gene selection (FGS)}
\begin{figure}[htpb]
    \centering
\includegraphics[width=0.9\textwidth]{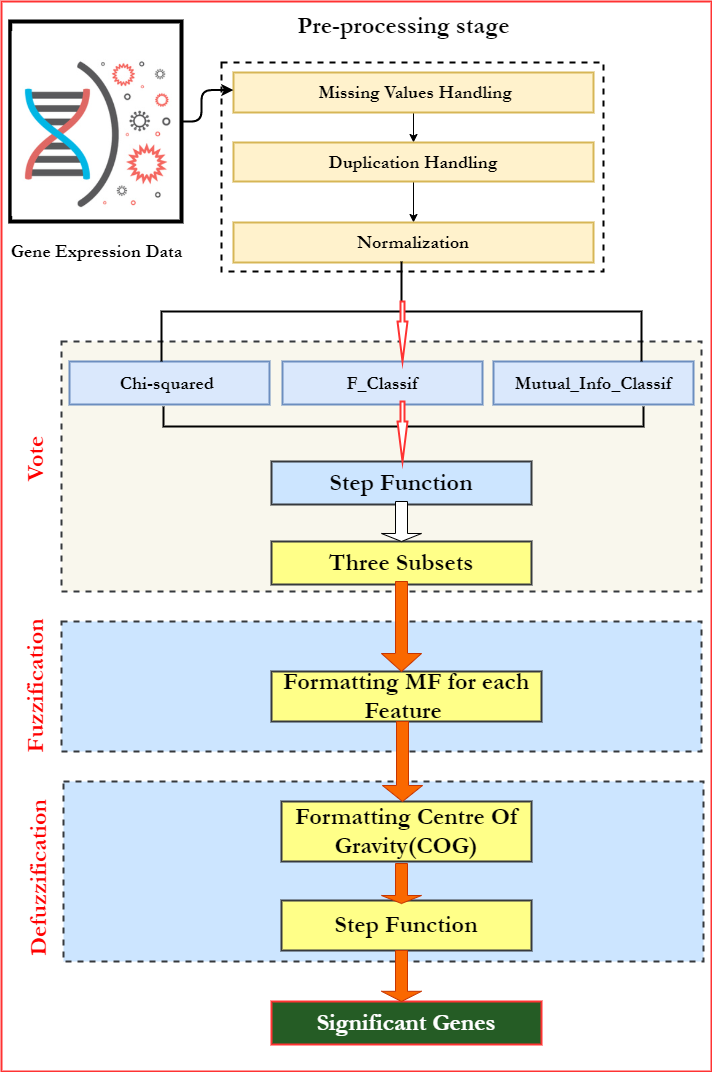}
\caption{Block Diagram of Proposed Fuzzy Gene selection Process}
    \label{fig2}
\end{figure}

 The proposed new fuzzy gene selection method of selecting the best subset of genes that were used as an identifier for the training classifier. The proposed FGS can be summarized in four major steps as shown in Figure\ref{fig2}. The steps are illustrated as follows:
\subsubsection{Pre-processing step}
The process of preparing raw data for use by machine learning algorithms is known as pre-possessing. Furthermore, it is the initial stage in data cleansing prior to analysis procedures such as feature selection or classification. The suggested algorithm employed three primary techniques of pre-processing, which are as follows:\\
1. Address the missing values: In general, missing values in a dataset have a negative influence on classifier performance, hence there are multiple ways for dealing with missing values (Eliminate Data Objects, Ignore the Missing Value During Analysis, and Estimate Missing Values). There are no missing values for a gene's expressed level in gene expression data. However, certain gene symbols are missing.
As a result, this stage removed only the raw data that does not contain the gene symbol.\\ 
2. Handle the duplication: simply eliminating the duplicated gene symbols.\\
3. Normalization is a procedure that is commonly used as part of data preparation for ML, particularly inside neural network classifier approaches. The primary goal of normalization is to modify the values of numeric columns in the dataset to use a similar scale without distorting variance in value ranges or losing information. The most common kind of normalization is min-max normalization, which was applied in this study. The normalization value is calculated using the equation below.. 
\begin{equation}
V =\frac{v-\mathrm{min}_{\mathrm{A}}} {{\mathrm{max}_{\mathrm{A}}}-{\mathrm{min}_{\mathrm{A}}}}
\label{g}
\end{equation}
Where:\\
maxA is the maximum value of original values for a feature.\\
minA is the minimum value of original values for a feature.\\
 and NmaxA,NminA are the maximum and minimum intervals of value.\\
V represents the feature value.
\subsubsection{Vote step}
Three feature selection approaches (MI,F-classif, and chi-squared) were used to select informative genes. Depending on the step function (SF), each feature selection approach chooses a different number of genes. The formula below has been used to compute the step function. This algorithm is intended to avoid using a limited number of selected genes, which may result in neglecting some genes with the same score when using a fixed number of genes, such as the top ten genes.  It is also worth noting that using this formula gives more flexibility to the step function value than using constant values such as 0.3. If non- or small-selected features by a feature selection method have scored equal to 0.3, we lose some essential features (genes) that could have been selected by other feature selection methods. 
\begin{equation}
SF=max(FSS)*0.3
\end{equation}
Where SF is step function, FSS is the feature selection score for all genes. \\
max is the maximum score for all features scored by the feature selection method.\\
The selected genes of this stage have scored either equal to the step function or  greater than the step function value that was calculated previously. 
\subsubsection{Fuzzification step}
This is the process of changing crisp data into fuzzy data using membership functions, with the goal of transforming the crisp data into data ranging between (0-1). There are different types of membership functions, the Triangular Membership Function was used in this work.\\
\begin{equation}
 Mf=\frac{\mathrm{W}_{\mathrm{i}}-a} {b-a}   
\end{equation}
Where MF is the membership function.\\
W is  the crisp value (score) for a gene.\\
a = lowest possible score (min).\\
b= highest possible score.\\
This membership function applied for the three feature selection methods which means, there are MF1, MF2, and MF3 in this work. 
\subsubsection{Defuzzification step}
This step is a process for converting the output data to crisp data. This step is the final stage of the gene selection method that has been used to select informative genes. The selected genes from these steps have been used as identifiers for training the classifier approaches.
\begin{equation}
 ASG=\frac{\mathrm{MF}_{\mathrm{i}}+\mathrm{MF}_{\mathrm{i}}+\mathrm{MF}_{\mathrm{i}}} {N} 
\end{equation}
Where ASG is the Average Score for a gene through the three feature selection methods.\\
MF is the membership function for each gene.
N is the number of feature selection methods that have been employed. In this work (N equal 3).\\
The two preceding phases show that different filter feature
selection approaches provide different scores for the same
gene. Fuzzification and Defuzzification were used to get a single score for each gene. As a result, as indicated in the equation below, using a step function for choosing the optimal subset of genes that would be used as identifiers for cancer classification.
\begin{equation}
SF =max(FSS)*0.5
\end{equation}
\subsection{Classifier Approaches }
\subsubsection{Support Vector Machine(SVM)}
It is applied for classification and regression challenges. However, SVM is typically applied to a classification problem because it accomplished outstanding performance in this area. SVM aims to create the best decision boundary (Hyperplane) to segregate the input data in different spaces. The SVM algorithm attempts to find the hyperplane in an n-dimensional space that segregates different data points [17][18]. 
Although, SVM has been widely used. However, it has some weaknesses. For example, SVM underperforms when the datasets are largely comparing it to small datasets. SVM is not  working well with datasets containing noise data for instance target classes are overlapping [19]. Additionally, it is not suited when the number of features is larger than the number of samples. These disadvantages of SVM have a high impact when applied to gene expression data because the gene expression data is noisy, and the number of genes is greater than the number of samples. 
\subsubsection{K-Nearest Neighbors (KNN)}
\begin{figure}[htpb]
    \centering
\includegraphics[width=0.9\textwidth]{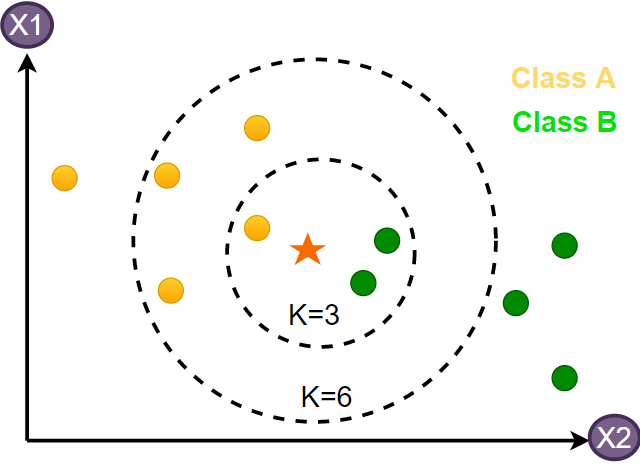}
\caption{KNN and its Hyperplane Selection}
    \label{fig4}
\end{figure}

It works on the assumption that similar things are positioned near to one another, making it more suitable for recommended system uses. To put it another way, KNN calculates the distance between the new point and the previously trained points (classes), so that the new point is predicted to the nearest distance of trained classes in feature space if it has two classes (Class A and Class B), as shown in Figure \ref{fig4}, and the "star" in red color represents the new class that requires prediction. Finding the best feature space (K) in KNN is critical because there is no standard method [18]. It often uses a large number of lists of integers to decide which one has the highest accuracy. As a consequence of this, the finest K will be picked. Although KNN is straightforward to use, there are several significant drawbacks. It is prone to noisy and missing data, is inefficient with large datasets, and contains data with high dimensionality.
 
\subsubsection{Decision Tree (DT)}
A decision tree is a supervised machine-learning technique that is used for both classification and regression challenges, however, it is mostly employed as a solution for classification purposes [18]. DT works under the  principle that the data is continuously split according to a certain parameter. It is easy to understand because it mimics the human process of making decisions and it requires less clean data compared with other ML approaches. However, it is complex compared with other algorithms because it consists of many layers and may have overfitting issues.It is also computationally  expensive as more class data labels are applied. The procedure of DT working can be concluded in five main steps as follows[21].\\  
1.Step1: DT starts with an entire dataset, assume S, in a node is called the root node. \\
2.Step2: Applying an attribute selection measure (ASM) to find the best attribute for given  a dataset. \\
3.Step3: Split the dataset into subsets that include the possible values for finding the best attribute for the given dataset. \\
4. Create the decision tress nodes, which have the best attribute.\\
5. Repeat step 3  partitioning the dataset into subsets for making a new decision tree, this process is continuously repeated until there is no possibility of classifying nodes namely leaf nodes that each leaf node presents one class or its probability [14].
\subsubsection{Gaussian Naive Bayes (GNB)}
Gaussian Naïve Bayes is supervised learning technique which relies on Bayes theorem that is employed for classification challenge and specifically for text classification because it is more suited to high dimensional training datasets [22]. It is considered one of the top 10 classifier techniques in data mining [23]. It is also characterized by faster prediction compared with other classifier models , easy to build and most effective in classification problems. However, GNB presumes that all features are independent which means it misses the possibility to learn the relationship between features [24][22]. Another drawback of GNB is hardly identifying the conditional independence in microarray data [25]. GNB works by taking each data point and assigning it to whichever class is nearest to it. It disguised not only calculating the distance by employing Euclidean distance between the new points and trained class, but it also calculates how this compares to the class variance. For each dimension, the z-score is calculated, and the distance from the mean is divided by the standard deviation [26]. 
\subsubsection{Multilayer Perceptron(MLP)}
MLP is a type of feedforward neural network (ANN) that is vastly used in pattern recognition, classification challenges, and prediction. It is mostly employed to solve supervised learning problems [17]. MLP maps the input to the output in a single direction of data and calculations. Generally, it consists of three perceptron or layers, an input layer, an output layer and at least one in between called a hidden layer[27]. Each layer in MLP is fully connected with the next layer. The input layer is used to receive the signal from the outside  world to the network, hidden layers perform the arithmetic operations from the input layer to the output layer while the output layer is responsible of making the decision(prediction). As a result, the output layer aims to transfer the information to the outside environment. Each layer in MLP is composed of a number of nodes (neurons). Most importantly, MLP work can be summarized in four main steps:\\
1) Step 1: propagating the input data forwarding from the  input layer to the output layer.\\ 
2)	Step 2:MLP is learned by updating the connection weights between the neurons to ensure a backpropagation algorithm is applied after input data of each node in MLP is processed[27]. \\
3)	Step 3:Calculate the errors by finding the difference between the predicted classes by MLP and the known classes and employ supervised learning to learn MLP to reduce the calculated errors.\\
4) The previous three steps will be repeated over multiple iterations to learn perfect weights. 
\subsection{Cross Validation}
\begin{figure}[htpb]
    \centering
\includegraphics[width=0.9\textwidth]{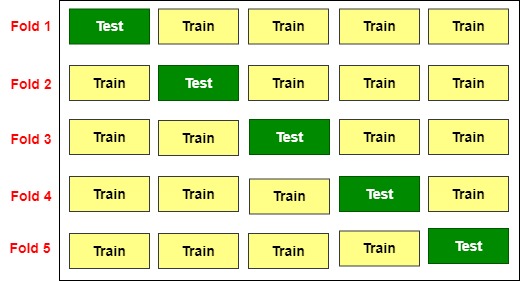}
     \caption{KFold Cross Validation Process with K=5}
    \label{fig5}
\end{figure}

Cross Validation in ML is a statistical method that aims to minimize or avoid overfitting issues in different classifier approaches. Rather than training a model on one training dataset, Cross Validation method allows training the model on many datasets. By splitting  the dataset into multiple folds and training the model on different folds [20]. As a result, the model achieves generalization capabilities which is a good sign of a robust model. It also assists to indicate a more accurate estimate of algorithm prediction performance.
The datasets split in kfold such as 5 as shown Figure\ref{fig5}. 
\subsection{Evaluation Measurement Methods}
This section is the evaluation tools that were used to evaluate the performance of the proposed model against the other previous models or compare the performance of classifier methods when the new fuzzy gene selection method was employed against the classifier methods when the fuzzy gene selection was not applied. As a result, these evaluation parameters are used for measuring the performance of a model. There are four evaluation measurements that must be explained to demonstrate that this proposed study outperformed the previous studies. The evaluation measurements are as follows:\\ 
Accuracy (AC) is an evaluation measurement that is utilized to determine which model is the best for a given dataset in AI. A ratio of correctly predicted observations to the total observations is called as accuracy in AI. The formula below is used to calculate it mathematically [28]:
\begin{equation}
Accuracy = \frac{TP + TN}{TP+ FP+ TN + FN}
\label{Eq:Accuracy}
\end{equation}
Where TP is True Positive, TN is True Negative, FP is False Positive and FN is False Negative.\\
A TP is the correctly predicted positive value which means that the value of the actual class is cancer and the value of the predicted class is also cancer.\\
A TN is an outcome where the model correctly predicts the negative class. A FP is an outcome where the model incorrectly predicts the positive class. FN is an outcome where the model incorrectly predicts the negative class .\\
Precision (Pre) is the ratio of correctly predicted positive observations to the total predicted positive observations as described in [30]
\begin{equation}
	Precision = \frac{TP }{TP + FP}
	\label{Eq:Precission}
\end{equation}
A recall (Rec) is the fraction of retrieved instances among all relevant instances. It is also known as sensitivity. The recall formula is illustrated as [28]:
\begin{equation}
       Recall = \frac{TP }{TP + FN}
	\label{Eq:Recall }
\end{equation}
The F1 score (F1) has combined the precision and recall of a classifier into a single metric by taking their harmonic mean, where a perfect F1 score has a value of 1 and the worst score at 0 [28]:
\begin{equation}
        F1 = 2\times \frac{precision \times recall}{precision + recall}
	\label{Eq:F1}
\end{equation}

\section{The proposed model}\label{sec4}
The proposed model may be divided into three basic stages of development. These phases were completed in the following order:\\  
1. The Pre-processing stage is prior to  machine learning included the removal of the raw data that had missing or duplicate gene symbols. The data were normalized by using a min-max normalization algorithm that aims to re-scale the data between (0-1).\\ 
2. The gene selection step, which was intended to select the optimal subset of informative genes that would be used as identifiers for training classifier algorithms, is the most significant stage of the proposed model. This stage can be represented by the following two points: To begin, we used three feature selection approaches (MI, F-classif, and chi-squared) with a step function to select a subset (the determined step function was displayed in the voting stage).
Second, the developed fuzzy gene selection approach employed fuzzy logic in a further analysis to choose fewer and more significant genes. The suggested FGS employed Triangular Membership Function fuzzification and center of gravity defuzzification with a step function (shown in the defuzzification phase) to choose informative ones with a strong influence on cancer classification.\\
3. Classifier stage: the proposed algorithm used Multi-layer Perceptron Classifier with three hidden layers. The output of the fuzzy gene selection method(selected genes) was used as an input layer for MLP (node number of input layer based on selected genes), three hidden layers were utilized (300,200,100 nodes) and one output layer which is the output of the classification(normal or malignant for binary classification and the class name for multiclasses datasets).\\
Summary: The total number of layers for the proposed model fifteen layers illustrated as follows: One input layer, three hidden layers for pre-processing stage (missing values, duplication, and normalization),three parallel hidden layers for filter feature selection methods. Two hidden layers for fuzzification (Triangular Membership Function) and defuzzification (Center of gravity). Three hidden layers for MLP classifier. Finally, one output layer. The number of input nodes is flexible which is based on the number features (number of genes) includes (the number of nodes when filter selection methods employed and the number of nodes when the fuzzy logic applied).

\begin{figure*}[tb] 
\centering
 \makebox[\textwidth]{\includegraphics[width=.9\paperwidth]{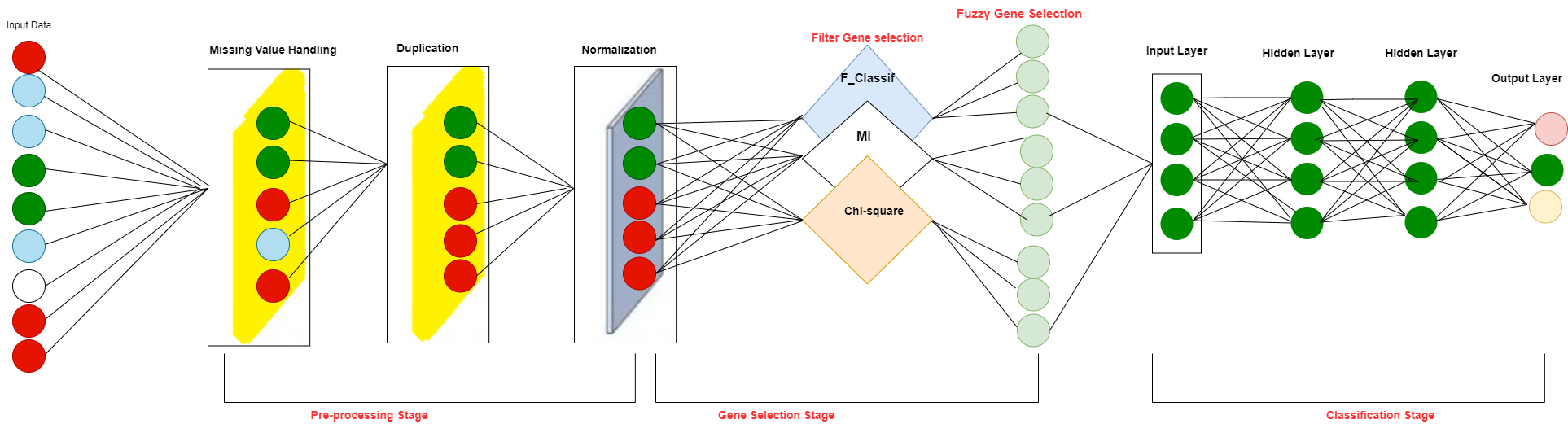}}
\caption{The Proposed Model Structure}
\label{fig:Fig9}
\end{figure*}
\section{Results }\label{sec10}
\subsection{Datasets used}
Six gene expression datasets of different types of cancer were used for training and testing the proposed model. The datasets comprised RNA-seq and Microarray tools were used to evaluate the proposed fuzzy gene selection algorithm with the two different measurement tools for measuring the expressed level of gene activity. The datasets were obtained from TCGA and GEO (GSE45827, GSE14520, GSE77314, GSE19804, TCGA, and GSE33630). The total number of samples from the six datasets was 3,011 for multi and binary classes more details were described in ( Table\ref{tab2}).  
To avoid overfitting in the training stage of the algorithm, the cross-validation method has been used with 5 Kfolds to split the datasets into multiple folds and train the algorithm on different folds.
In Table \ref{tab2}, KIRC stands for Kidney renal cell cancer, LUAD stands for Lung adenocarcinoma, LUSC stands for Lung squamous cell carcinoma, and UCEC is for Uterine corpus endometrial carcinoma.\\ 
\begin{table*}[htpb]
\caption{Summary of Datasets were Employed for Training and Testing The Proposed Model}
\centering
\label{tab2}
\begin{tabular}{|m{5em}|m{5em}|m{8em}|m{3em}|m{4em}|m{2em}|m{4em}|}\\
\hline
Dataset &  Tools &
  N-samples &
   N-Genes&
  Cancer Types &
  N-Class &
  Reference \\ \hline
  GSE45827 &
  Microarray  &
  155 (Basal        41, Her2  30,
  Luminal B  30,
Luminal A    29,
CellLine     14,
Normal       11
) &
 29873 &
  Breast cancer subtypes &
6 &
  {[}11{]} \\ \hline
   GSE14520 &
   Microarray   &
445 ( Cancer 227, Normal 218) &
  13425 &
  Liver Cancer &
    2 &
  {[}11{]} \\ \hline
  GSE77314 &
   RNA-seq  &  100 (Cancer 50, Normal 50)&
  29087&
   Liver Cancer  &
    2 &
  {[}11{]} \\ \hline
GSE19804&
   Microarray &
  120 (Cancer 60, Normal 60) &
  45782&
   Lung Cancer  &
    2 &
  {[}11{]} \\ \hline
  TCGA &
   RNA-seq  &
  2086 ( BRCA   878,
KIRC       537, UCEC   269,
 LUSC     240,LUAD     162)
 &
972 &
BRCA, KIRC, LUAD, LUSC, UCEC  &
    5&
  {[}29{]} \\ \hline
 GSE33630 &
  Microarray  &
  105 (PTC 49, Normal    45, ATC    11) &
  23518 &
   Thyroid &
    3 &
  {[}11{]} \\ \hline
\end{tabular}\\
\end{table*}
\subsection{Obtained results}
This section investigates the usage of six datasets across five classifier approaches, comparing the use of a fuzzy gene selection method and demonstrating the benefits of using the suggested fuzzy gene selection methodology. In this paper, we examine how FGS affects the performance of cancer classification models. The full details are presented (Table \ref{tab3} and Table \ref{tab4})  of the datasets used for training and testing the models, cancer types, and the achieved accuracy, precision, recall, and f1-score  before the fuzzy gene selection method was applied and after the fuzzy gene selection method was used. 
\begin{table*}[htpb]
\caption{Comparing five classifier approaches  when applying and omitting FGS}
\centering
\label{tab3}
\begin{tabular}{|c|c|c|c|c|c|c|c|c|}
\hline
Dataset &
  Class Types &
  FS method &
  N-Genes &
  Classifier &
  Ac&
  Pre &
  Rec&
  F1
  \\ \hline
  \begin{tabular}[c]{@{}l@{}}\end{tabular} &
  \begin{tabular}[c]{@{}l@{}}\end{tabular} &
  \begin{tabular}[c]{@{}l@{}}\end{tabular} &
  \begin{tabular}[c]{@{}l@{}} \end{tabular} &
   \begin{tabular}[c]{@{}l@{}} DT.\end{tabular} &
    \begin{tabular}[c]{@{}l@{}} 90\% \end{tabular}&
    \begin{tabular}[c]{@{}l@{}} 90.6\% \end{tabular}&
    \begin{tabular}[c]{@{}l@{}} 88.9\% \end{tabular}&
    \begin{tabular}[c]{@{}l@{}} 89.7\% \end{tabular}\\
    \begin{tabular}[c]{@{}l@{}}\end{tabular} &
  \begin{tabular}[c]{@{}l@{}} \end{tabular} &
  \begin{tabular}[c]{@{}l@{}}\end{tabular} &
  \begin{tabular}[c]{@{}l@{}} \end{tabular} &
   \begin{tabular}[c]{@{}l@{}} KNN.\end{tabular} &
    \begin{tabular}[c]{@{}l@{}} 94\% \end{tabular}&
    \begin{tabular}[c]{@{}l@{}} 91\% \end{tabular}&
    \begin{tabular}[c]{@{}l@{}} 97.6\% \end{tabular}&
    \begin{tabular}[c]{@{}l@{}} 94\% \end{tabular}\\
    
     \begin{tabular}[c]{@{}l@{}}GSE14520\end{tabular} &
  \begin{tabular}[c]{@{}l@{}} Binary class\end{tabular} &
  \begin{tabular}[c]{@{}l@{}}No\end{tabular} &
  \begin{tabular}[c]{@{}l@{}}13425 \end{tabular} &
   \begin{tabular}[c]{@{}l@{}} SVM.\end{tabular} &
    \begin{tabular}[c]{@{}l@{}} 97\% \end{tabular}&
    \begin{tabular}[c]{@{}l@{}} 96\% \end{tabular}&
    \begin{tabular}[c]{@{}l@{}} 97.6\% \end{tabular}&
    \begin{tabular}[c]{@{}l@{}} 97\% \end{tabular}\\
     \begin{tabular}[c]{@{}l@{}}\end{tabular} &
  \begin{tabular}[c]{@{}l@{}} \end{tabular} &
  \begin{tabular}[c]{@{}l@{}}\end{tabular} &
  \begin{tabular}[c]{@{}l@{}} \end{tabular} &
   \begin{tabular}[c]{@{}l@{}} GNB.\end{tabular} &
    \begin{tabular}[c]{@{}l@{}} 95\% \end{tabular}&
    \begin{tabular}[c]{@{}l@{}} 95.6\% \end{tabular}&
    \begin{tabular}[c]{@{}l@{}} 94\% \end{tabular}&
    \begin{tabular}[c]{@{}l@{}} 94.8\% \end{tabular}\\
    
 \begin{tabular}[c]{@{}l@{}}\end{tabular} &
  \begin{tabular}[c]{@{}l@{}} \end{tabular} &
  \begin{tabular}[c]{@{}l@{}}\end{tabular} &
  \begin{tabular}[c]{@{}l@{}} \end{tabular} &
   \begin{tabular}[c]{@{}l@{}} MLP.\end{tabular} &
    \begin{tabular}[c]{@{}l@{}} 86.7\% \end{tabular}&
     \begin{tabular}[c]{@{}l@{}} 76.5\% \end{tabular}&
      \begin{tabular}[c]{@{}l@{}} 76.7\% \end{tabular}&
       \begin{tabular}[c]{@{}l@{}} 76.5\% \end{tabular}\\ \hline
     \begin{tabular}[c]{@{}l@{}}\end{tabular} &
  \begin{tabular}[c]{@{}l@{}}\end{tabular} &
  \begin{tabular}[c]{@{}l@{}}\end{tabular} &
  \begin{tabular}[c]{@{}l@{}} \end{tabular} &
   \begin{tabular}[c]{@{}l@{}} DT.\end{tabular} &
    \begin{tabular}[c]{@{}l@{}} 96\% \end{tabular}&
     \begin{tabular}[c]{@{}l@{}} 95\% \end{tabular}&
      \begin{tabular}[c]{@{}l@{}} 97\% \end{tabular}&
       \begin{tabular}[c]{@{}l@{}} 96\% \end{tabular}\\
    \begin{tabular}[c]{@{}l@{}}\end{tabular} &
  \begin{tabular}[c]{@{}l@{}} \end{tabular} &
  \begin{tabular}[c]{@{}l@{}}\end{tabular} &
  \begin{tabular}[c]{@{}l@{}} \end{tabular} &
   \begin{tabular}[c]{@{}l@{}} KNN.\end{tabular} &
    \begin{tabular}[c]{@{}l@{}} 96.6\% \end{tabular}&
    \begin{tabular}[c]{@{}l@{}} 96\% \end{tabular}&
    \begin{tabular}[c]{@{}l@{}} 97\% \end{tabular}&
    \begin{tabular}[c]{@{}l@{}} 96.6\% \end{tabular}\\
     \begin{tabular}[c]{@{}l@{}}GSE14520\end{tabular} &
  \begin{tabular}[c]{@{}l@{}} Binary class\end{tabular} &
  \begin{tabular}[c]{@{}l@{}}FGS \end{tabular} &
  \begin{tabular}[c]{@{}l@{}}23 \end{tabular} &
   \begin{tabular}[c]{@{}l@{}} SVM.\end{tabular} &
    \begin{tabular}[c]{@{}l@{}} 96\% \end{tabular}&
    \begin{tabular}[c]{@{}l@{}} 95.6\% \end{tabular}&
    \begin{tabular}[c]{@{}l@{}} 96\% \end{tabular}&
    \begin{tabular}[c]{@{}l@{}} 96\% \end{tabular}\\
     \begin{tabular}[c]{@{}l@{}}\end{tabular} &
  \begin{tabular}[c]{@{}l@{}} \end{tabular} &
  \begin{tabular}[c]{@{}l@{}}\end{tabular} &
  \begin{tabular}[c]{@{}l@{}} \end{tabular} &
   \begin{tabular}[c]{@{}l@{}} GNB.\end{tabular} &
    \begin{tabular}[c]{@{}l@{}} 96.6\% \end{tabular}&
    \begin{tabular}[c]{@{}l@{}} 96\% \end{tabular}&
    \begin{tabular}[c]{@{}l@{}} 97\% \end{tabular}&
    \begin{tabular}[c]{@{}l@{}} 96.6\% \end{tabular}\\
    
 \begin{tabular}[c]{@{}l@{}}\end{tabular} &
  \begin{tabular}[c]{@{}l@{}} \end{tabular} &
  \begin{tabular}[c]{@{}l@{}}\end{tabular} &
  \begin{tabular}[c]{@{}l@{}} \end{tabular} &
   \begin{tabular}[c]{@{}l@{}} MLP.\end{tabular} &
    \begin{tabular}[c]{@{}l@{}} 96\% \end{tabular}&
    \begin{tabular}[c]{@{}l@{}} 96\% \end{tabular}&
    \begin{tabular}[c]{@{}l@{}} 96\% \end{tabular}&
    \begin{tabular}[c]{@{}l@{}} 96\% \end{tabular}\\ \hline
  \begin{tabular}[c]{@{}l@{}}\end{tabular} &
  \begin{tabular}[c]{@{}l@{}}\end{tabular} &
  \begin{tabular}[c]{@{}l@{}}\end{tabular} &
  \begin{tabular}[c]{@{}l@{}} \end{tabular} &
   \begin{tabular}[c]{@{}l@{}} DT.\end{tabular} &
    \begin{tabular}[c]{@{}l@{}} 87.6\% \end{tabular}&
    \begin{tabular}[c]{@{}l@{}} 77.6\% \end{tabular}&\
    \begin{tabular}[c]{@{}l@{}} 81\% \end{tabular}&\
    \begin{tabular}[c]{@{}l@{}} 79\% \end{tabular}\\
    \begin{tabular}[c]{@{}l@{}}\end{tabular} &
  \begin{tabular}[c]{@{}l@{}} \end{tabular} &
  \begin{tabular}[c]{@{}l@{}}\end{tabular} &
  \begin{tabular}[c]{@{}l@{}} \end{tabular} &
   \begin{tabular}[c]{@{}l@{}} KNN.\end{tabular} &
    \begin{tabular}[c]{@{}l@{}} 91\% \end{tabular}&
    \begin{tabular}[c]{@{}l@{}} 87.7\% \end{tabular}&
    \begin{tabular}[c]{@{}l@{}} 86.5\% \end{tabular}&
    \begin{tabular}[c]{@{}l@{}} 86\% \end{tabular}\\
     \begin{tabular}[c]{@{}l@{}}GSE33630\end{tabular} &
  \begin{tabular}[c]{@{}l@{}} Multiclass \end{tabular} &
  \begin{tabular}[c]{@{}l@{}}No\end{tabular} &
  \begin{tabular}[c]{@{}l@{}}23516 \end{tabular} &
   \begin{tabular}[c]{@{}l@{}} SVM.\end{tabular} &
    \begin{tabular}[c]{@{}l@{}}93\% \end{tabular}&
    \begin{tabular}[c]{@{}l@{}} 95\% \end{tabular}&
    \begin{tabular}[c]{@{}l@{}} 92\% \end{tabular}&
    \begin{tabular}[c]{@{}l@{}} 92\% \end{tabular}\\
     \begin{tabular}[c]{@{}l@{}}\end{tabular} &
  \begin{tabular}[c]{@{}l@{}} \end{tabular} &
  \begin{tabular}[c]{@{}l@{}}\end{tabular} &
  \begin{tabular}[c]{@{}l@{}} \end{tabular} &
   \begin{tabular}[c]{@{}l@{}} GNB.\end{tabular} &
    \begin{tabular}[c]{@{}l@{}} 90\% \end{tabular}&
     \begin{tabular}[c]{@{}l@{}} 93.7\% \end{tabular}&
      \begin{tabular}[c]{@{}l@{}} 89.7\% \end{tabular}&
       \begin{tabular}[c]{@{}l@{}} 90\% \end{tabular}\\
    
 \begin{tabular}[c]{@{}l@{}}\end{tabular} &
  \begin{tabular}[c]{@{}l@{}} \end{tabular} &
  \begin{tabular}[c]{@{}l@{}}\end{tabular} &
  \begin{tabular}[c]{@{}l@{}} \end{tabular} &
   \begin{tabular}[c]{@{}l@{}} MLP.\end{tabular} &
    \begin{tabular}[c]{@{}l@{}} 72\% \end{tabular}&
    \begin{tabular}[c]{@{}l@{}} 55.6\% \end{tabular}&
    \begin{tabular}[c]{@{}l@{}} 64.5\% \end{tabular}&
    \begin{tabular}[c]{@{}l@{}} 58.5\% \end{tabular}\\\hline
   
     \begin{tabular}[c]{@{}l@{}}\end{tabular} &
  \begin{tabular}[c]{@{}l@{}}\end{tabular} &
  \begin{tabular}[c]{@{}l@{}}\end{tabular} &
  \begin{tabular}[c]{@{}l@{}} \end{tabular} &
   \begin{tabular}[c]{@{}l@{}} DT.\end{tabular} &
    \begin{tabular}[c]{@{}l@{}} 93\% \end{tabular}&
    \begin{tabular}[c]{@{}l@{}} 93\% \end{tabular}&
    \begin{tabular}[c]{@{}l@{}} 93.5\% \end{tabular}&
    \begin{tabular}[c]{@{}l@{}} 92.5\% \end{tabular}\\
    \begin{tabular}[c]{@{}l@{}}\end{tabular} &
  \begin{tabular}[c]{@{}l@{}} \end{tabular} &
  \begin{tabular}[c]{@{}l@{}}\end{tabular} &
  \begin{tabular}[c]{@{}l@{}} \end{tabular} &
   \begin{tabular}[c]{@{}l@{}} KNN.\end{tabular} &
    \begin{tabular}[c]{@{}l@{}} 94\% \end{tabular}&
     \begin{tabular}[c]{@{}l@{}} 96\% \end{tabular}&
    \begin{tabular}[c]{@{}l@{}} 92.8\% \end{tabular}&
     \begin{tabular}[c]{@{}l@{}} 93\% \end{tabular}\\
    
     \begin{tabular}[c]{@{}l@{}}GSE33630\end{tabular} &
  \begin{tabular}[c]{@{}l@{}} Multiclass \end{tabular} &
  \begin{tabular}[c]{@{}l@{}}FGS \end{tabular} &
  \begin{tabular}[c]{@{}l@{}}76 \end{tabular} &
   \begin{tabular}[c]{@{}l@{}} SVM.\end{tabular} &
    \begin{tabular}[c]{@{}l@{}} 94\% \end{tabular}&
      \begin{tabular}[c]{@{}l@{}} 96\% \end{tabular}&
        \begin{tabular}[c]{@{}l@{}} 92.8\% \end{tabular}&
          \begin{tabular}[c]{@{}l@{}} 93\% \end{tabular}\\
     \begin{tabular}[c]{@{}l@{}}\end{tabular} &
  \begin{tabular}[c]{@{}l@{}} \end{tabular} &
  \begin{tabular}[c]{@{}l@{}}\end{tabular} &
  \begin{tabular}[c]{@{}l@{}} \end{tabular} &
   \begin{tabular}[c]{@{}l@{}} GNB.\end{tabular} &
    \begin{tabular}[c]{@{}l@{}} 92\% \end{tabular}&
     \begin{tabular}[c]{@{}l@{}} 88\% \end{tabular}&
      \begin{tabular}[c]{@{}l@{}} 99.8\% \end{tabular}&
       \begin{tabular}[c]{@{}l@{}} 88.8\% \end{tabular}\\
 \begin{tabular}[c]{@{}l@{}}\end{tabular} &
  \begin{tabular}[c]{@{}l@{}} \end{tabular} &
  \begin{tabular}[c]{@{}l@{}}\end{tabular} &
  \begin{tabular}[c]{@{}l@{}} \end{tabular} &
   \begin{tabular}[c]{@{}l@{}} MLP.\end{tabular} &
    \begin{tabular}[c]{@{}l@{}} 93\% \end{tabular}&
     \begin{tabular}[c]{@{}l@{}} 95\% \end{tabular}&
      \begin{tabular}[c]{@{}l@{}} 92\% \end{tabular}&
       \begin{tabular}[c]{@{}l@{}} 92.5\% \end{tabular}\\ \hline
 \begin{tabular}[c]{@{}l@{}}\end{tabular} &
  \begin{tabular}[c]{@{}l@{}}\end{tabular} &
  \begin{tabular}[c]{@{}l@{}}\end{tabular} &
  \begin{tabular}[c]{@{}l@{}} \end{tabular} &
   \begin{tabular}[c]{@{}l@{}} DT.\end{tabular} &
    \begin{tabular}[c]{@{}l@{}} 91\% \end{tabular}
    &
     \begin{tabular}[c]{@{}l@{}} 87\% \end{tabular}&
      \begin{tabular}[c]{@{}l@{}} 85\% \end{tabular}&
     \begin{tabular}[c]{@{}l@{}} 85.8\% \end{tabular}\\
    \begin{tabular}[c]{@{}l@{}}\end{tabular} &
  \begin{tabular}[c]{@{}l@{}} \end{tabular} &
  \begin{tabular}[c]{@{}l@{}}\end{tabular} &
  \begin{tabular}[c]{@{}l@{}} \end{tabular} &
   \begin{tabular}[c]{@{}l@{}} KNN.\end{tabular} &
    \begin{tabular}[c]{@{}l@{}} 88\% \end{tabular}&
    \begin{tabular}[c]{@{}l@{}} 83\% \end{tabular}&
    \begin{tabular}[c]{@{}l@{}} 81.5\% \end{tabular}&
    \begin{tabular}[c]{@{}l@{}} 81.9\% \end{tabular}\\
    
     \begin{tabular}[c]{@{}l@{}}TCGA \end{tabular} &
  \begin{tabular}[c]{@{}l@{}} Multiclass \end{tabular} &
  \begin{tabular}[c]{@{}l@{}}No\end{tabular} &
  \begin{tabular}[c]{@{}l@{}}971 \end{tabular} &
   \begin{tabular}[c]{@{}l@{}} SVM.\end{tabular} &
    \begin{tabular}[c]{@{}l@{}} 95\% \end{tabular}&
     \begin{tabular}[c]{@{}l@{}} 91.6\% \end{tabular}&
      \begin{tabular}[c]{@{}l@{}} 91.8\% \end{tabular}&
       \begin{tabular}[c]{@{}l@{}} 91.6\% \end{tabular}\\
     \begin{tabular}[c]{@{}l@{}}\end{tabular} &
  \begin{tabular}[c]{@{}l@{}} \end{tabular} &
  \begin{tabular}[c]{@{}l@{}}\end{tabular} &
  \begin{tabular}[c]{@{}l@{}} \end{tabular} &
   \begin{tabular}[c]{@{}l@{}} GNB.\end{tabular} &
    \begin{tabular}[c]{@{}l@{}} 94\% \end{tabular}&
     \begin{tabular}[c]{@{}l@{}} 89.7\% \end{tabular}&
      \begin{tabular}[c]{@{}l@{}} 92\% \end{tabular}&
       \begin{tabular}[c]{@{}l@{}} 90.7\% \end{tabular}\\
 \begin{tabular}[c]{@{}l@{}}\end{tabular} &
  \begin{tabular}[c]{@{}l@{}} \end{tabular} &
  \begin{tabular}[c]{@{}l@{}}\end{tabular} &
  \begin{tabular}[c]{@{}l@{}} \end{tabular} &
   \begin{tabular}[c]{@{}l@{}} MLP.\end{tabular} &
    \begin{tabular}[c]{@{}l@{}} 94\% \end{tabular}&
    \begin{tabular}[c]{@{}l@{}} 90.8\% \end{tabular}&
    \begin{tabular}[c]{@{}l@{}} 89.8\% \end{tabular}&
    \begin{tabular}[c]{@{}l@{}} 90\% \end{tabular}\\ \hline
   
     \begin{tabular}[c]{@{}l@{}}\end{tabular} &
  \begin{tabular}[c]{@{}l@{}}\end{tabular} &
  \begin{tabular}[c]{@{}l@{}}\end{tabular} &
  \begin{tabular}[c]{@{}l@{}} \end{tabular} &
   \begin{tabular}[c]{@{}l@{}} DT.\end{tabular} &
    \begin{tabular}[c]{@{}l@{}} 91.7\% \end{tabular}&
    \begin{tabular}[c]{@{}l@{}} 88\% \end{tabular}&
    \begin{tabular}[c]{@{}l@{}} 87\% \end{tabular}&
    \begin{tabular}[c]{@{}l@{}} 86.5\% \end{tabular}\\
    \begin{tabular}[c]{@{}l@{}}\end{tabular} &
  \begin{tabular}[c]{@{}l@{}} \end{tabular} &
  \begin{tabular}[c]{@{}l@{}}\end{tabular} &
  \begin{tabular}[c]{@{}l@{}} \end{tabular} &
   \begin{tabular}[c]{@{}l@{}} KNN.\end{tabular} &
    \begin{tabular}[c]{@{}l@{}} 93.6\% \end{tabular}&
    \begin{tabular}[c]{@{}l@{}} 89.8\% \end{tabular}&
    \begin{tabular}[c]{@{}l@{}} 90\% \end{tabular}&
    \begin{tabular}[c]{@{}l@{}} 89.6\% \end{tabular}\\
     \begin{tabular}[c]{@{}l@{}}TCGA \end{tabular} &
  \begin{tabular}[c]{@{}l@{}} Multiclass \end{tabular} &
  \begin{tabular}[c]{@{}l@{}}FGS \end{tabular} &
  \begin{tabular}[c]{@{}l@{}}25 \end{tabular} &
   \begin{tabular}[c]{@{}l@{}} SVM.\end{tabular} &
    \begin{tabular}[c]{@{}l@{}} 94	\% \end{tabular}&
    \begin{tabular}[c]{@{}l@{}} 90.5\% \end{tabular}&
    \begin{tabular}[c]{@{}l@{}} 90.7\% \end{tabular}&
    \begin{tabular}[c]{@{}l@{}}90.5\% \end{tabular}\\
     \begin{tabular}[c]{@{}l@{}}\end{tabular} &
  \begin{tabular}[c]{@{}l@{}} \end{tabular} &
  \begin{tabular}[c]{@{}l@{}}\end{tabular} &
  \begin{tabular}[c]{@{}l@{}} \end{tabular} &
   \begin{tabular}[c]{@{}l@{}} GNB.\end{tabular} &
    \begin{tabular}[c]{@{}l@{}} 92\% \end{tabular}&
    \begin{tabular}[c]{@{}l@{}} 87.7\% \end{tabular}&
    \begin{tabular}[c]{@{}l@{}} 90.8\% \end{tabular}&
    \begin{tabular}[c]{@{}l@{}} 89\% \end{tabular}\\
    \begin{tabular}[c]{@{}l@{}}\end{tabular} &
  \begin{tabular}[c]{@{}l@{}} \end{tabular} &
  \begin{tabular}[c]{@{}l@{}}\end{tabular} &
  \begin{tabular}[c]{@{}l@{}} \end{tabular} &
   \begin{tabular}[c]{@{}l@{}} MLP.\end{tabular} &
    \begin{tabular}[c]{@{}l@{}} 95\% \end{tabular}&
    \begin{tabular}[c]{@{}l@{}} 92\% \end{tabular}&
    \begin{tabular}[c]{@{}l@{}} 91.6\% \end{tabular}&
    \begin{tabular}[c]{@{}l@{}} 91.6\% \end{tabular}\\
    \hline
  \end{tabular}
\end{table*}\\
\begin{table*}[htpb]
\caption{Comparing five classifier approaches  when applying and omitting FGS}
\centering
\label{tab4}
\begin{tabular}{|c|c|c|c|c|c|c|c|c|}
\hline
Dataset &
  Class Types &
  FS method &
  N-Genes &
  Classifier &
  Ac&
  Pre &
  Rec&
  F1
  \\ \hline
 \begin{tabular}[c]{@{}l@{}}\end{tabular} &
  \begin{tabular}[c]{@{}l@{}}\end{tabular} &
  \begin{tabular}[c]{@{}l@{}}\end{tabular} &
  \begin{tabular}[c]{@{}l@{}} \end{tabular} &
   \begin{tabular}[c]{@{}l@{}} DT.\end{tabular} &
    \begin{tabular}[c]{@{}l@{}} 89\% \end{tabular}&
    \begin{tabular}[c]{@{}l@{}} 90\% \end{tabular}&
    \begin{tabular}[c]{@{}l@{}} 88\% \end{tabular}&
    \begin{tabular}[c]{@{}l@{}} 90\% \end{tabular}\\
    \begin{tabular}[c]{@{}l@{}}\end{tabular} &
  \begin{tabular}[c]{@{}l@{}} \end{tabular} &
  \begin{tabular}[c]{@{}l@{}}\end{tabular} &
  \begin{tabular}[c]{@{}l@{}} \end{tabular} &
   \begin{tabular}[c]{@{}l@{}} KNN.\end{tabular} &
    \begin{tabular}[c]{@{}l@{}} 90.8\% \end{tabular}&
    \begin{tabular}[c]{@{}l@{}} 88\% \end{tabular}&
    \begin{tabular}[c]{@{}l@{}} 95\% \end{tabular}&
    \begin{tabular}[c]{@{}l@{}} 91\% \end{tabular}\\
     \begin{tabular}[c]{@{}l@{}}GSE19804 \end{tabular} &
  \begin{tabular}[c]{@{}l@{}} Binary class \end{tabular} &
  \begin{tabular}[c]{@{}l@{}}No\end{tabular} &
  \begin{tabular}[c]{@{}l@{}}45782 \end{tabular} &
   \begin{tabular}[c]{@{}l@{}} SVM.\end{tabular} &
    \begin{tabular}[c]{@{}l@{}} 95.8\% \end{tabular}&
     \begin{tabular}[c]{@{}l@{}} 96.6\% \end{tabular}&
      \begin{tabular}[c]{@{}l@{}} 95\% \end{tabular}&
       \begin{tabular}[c]{@{}l@{}} 95.7\% \end{tabular}\\
     \begin{tabular}[c]{@{}l@{}}\end{tabular} &
  \begin{tabular}[c]{@{}l@{}} \end{tabular} &
  \begin{tabular}[c]{@{}l@{}}\end{tabular} &
  \begin{tabular}[c]{@{}l@{}} \end{tabular} &
   \begin{tabular}[c]{@{}l@{}} GNB.\end{tabular} &
    \begin{tabular}[c]{@{}l@{}} 92.5\% \end{tabular}&
      \begin{tabular}[c]{@{}l@{}} 95\% \end{tabular}&
        \begin{tabular}[c]{@{}l@{}} 90\% \end{tabular}&
          \begin{tabular}[c]{@{}l@{}} 91.9\% \end{tabular}\\
    
 \begin{tabular}[c]{@{}l@{}}\end{tabular} &
  \begin{tabular}[c]{@{}l@{}} \end{tabular} &
  \begin{tabular}[c]{@{}l@{}}\end{tabular} &
  \begin{tabular}[c]{@{}l@{}} \end{tabular} &
   \begin{tabular}[c]{@{}l@{}} MLP.\end{tabular} &
    \begin{tabular}[c]{@{}l@{}} 50\% \end{tabular}&
    \begin{tabular}[c]{@{}l@{}} 20\% \end{tabular}&
    \begin{tabular}[c]{@{}l@{}} 40\% \end{tabular}&
    \begin{tabular}[c]{@{}l@{}} 26.6\% \end{tabular} \\ \hline
   
     \begin{tabular}[c]{@{}l@{}}\end{tabular} &
  \begin{tabular}[c]{@{}l@{}}\end{tabular} &
  \begin{tabular}[c]{@{}l@{}}\end{tabular} &
  \begin{tabular}[c]{@{}l@{}} \end{tabular} &
   \begin{tabular}[c]{@{}l@{}} DT.\end{tabular} &
    \begin{tabular}[c]{@{}l@{}} 92.5\% \end{tabular}&
    \begin{tabular}[c]{@{}l@{}} 93.6\% \end{tabular}&
    \begin{tabular}[c]{@{}l@{}} 91.6\% \end{tabular}&
    \begin{tabular}[c]{@{}l@{}} 92\% \end{tabular}\\
    \begin{tabular}[c]{@{}l@{}}\end{tabular} &
  \begin{tabular}[c]{@{}l@{}} \end{tabular} &
  \begin{tabular}[c]{@{}l@{}}\end{tabular} &
  \begin{tabular}[c]{@{}l@{}} \end{tabular} &
   \begin{tabular}[c]{@{}l@{}} KNN.\end{tabular} &
    \begin{tabular}[c]{@{}l@{}} 96.6\% \end{tabular}&
     \begin{tabular}[c]{@{}l@{}} 96.7\% \end{tabular}&
      \begin{tabular}[c]{@{}l@{}} 96.6\% \end{tabular}&
       \begin{tabular}[c]{@{}l@{}} 96.6\% \end{tabular}\\
    
     \begin{tabular}[c]{@{}l@{}}GSE19804 \end{tabular} &
  \begin{tabular}[c]{@{}l@{}} Binary class  \end{tabular} &
  \begin{tabular}[c]{@{}l@{}}FGS \end{tabular} &
  \begin{tabular}[c]{@{}l@{}}36 \end{tabular} &
   \begin{tabular}[c]{@{}l@{}} SVM.\end{tabular} &
    \begin{tabular}[c]{@{}l@{}} 96.6\% \end{tabular}&
    \begin{tabular}[c]{@{}l@{}} 97\% \end{tabular}&
    \begin{tabular}[c]{@{}l@{}} 96.6\% \end{tabular}&
    \begin{tabular}[c]{@{}l@{}} 96.6\% \end{tabular}\\
     \begin{tabular}[c]{@{}l@{}}\end{tabular} &
  \begin{tabular}[c]{@{}l@{}} \end{tabular} &
  \begin{tabular}[c]{@{}l@{}}\end{tabular} &
  \begin{tabular}[c]{@{}l@{}} \end{tabular} &
   \begin{tabular}[c]{@{}l@{}} GNB.\end{tabular} &
    \begin{tabular}[c]{@{}l@{}} 95.8\% \end{tabular}&
    \begin{tabular}[c]{@{}l@{}} 96.7\% \end{tabular}&
    \begin{tabular}[c]{@{}l@{}} 95\% \end{tabular}&
    \begin{tabular}[c]{@{}l@{}} 95.7\% \end{tabular}\\
    
 \begin{tabular}[c]{@{}l@{}}\end{tabular} &
  \begin{tabular}[c]{@{}l@{}} \end{tabular} &
  \begin{tabular}[c]{@{}l@{}}\end{tabular} &
  \begin{tabular}[c]{@{}l@{}} \end{tabular} &
   \begin{tabular}[c]{@{}l@{}} MLP.\end{tabular} &
    \begin{tabular}[c]{@{}l@{}} 97.5\% \end{tabular}&
    \begin{tabular}[c]{@{}l@{}} 97\% \end{tabular}&
    \begin{tabular}[c]{@{}l@{}} 98\% \end{tabular}&
    \begin{tabular}[c]{@{}l@{}} 97.5\% \end{tabular}\\ \hline
 
    \begin{tabular}[c]{@{}l@{}}\end{tabular} &
  \begin{tabular}[c]{@{}l@{}}\end{tabular} &
  \begin{tabular}[c]{@{}l@{}}\end{tabular} &
  \begin{tabular}[c]{@{}l@{}} \end{tabular} &
   \begin{tabular}[c]{@{}l@{}} DT.\end{tabular} &
    \begin{tabular}[c]{@{}l@{}} 95\% \end{tabular}&
    \begin{tabular}[c]{@{}l@{}} 98\% \end{tabular}&
    \begin{tabular}[c]{@{}l@{}} 91.9\% \end{tabular}&
    \begin{tabular}[c]{@{}l@{}} 94\% \end{tabular}\\
    \begin{tabular}[c]{@{}l@{}}\end{tabular} &
  \begin{tabular}[c]{@{}l@{}} \end{tabular} &
  \begin{tabular}[c]{@{}l@{}}\end{tabular} &
  \begin{tabular}[c]{@{}l@{}} \end{tabular} &
   \begin{tabular}[c]{@{}l@{}} KNN.\end{tabular} &
    \begin{tabular}[c]{@{}l@{}} 88.9\% \end{tabular}&
      \begin{tabular}[c]{@{}l@{}} 82\% \end{tabular}&
        \begin{tabular}[c]{@{}l@{}} 100\% \end{tabular}&
          \begin{tabular}[c]{@{}l@{}} 90\% \end{tabular}\\
     \begin{tabular}[c]{@{}l@{}}GSE77314 \end{tabular} &
  \begin{tabular}[c]{@{}l@{}} Binary class \end{tabular} &
  \begin{tabular}[c]{@{}l@{}}No\end{tabular} &
  \begin{tabular}[c]{@{}l@{}}29087 \end{tabular} &
   \begin{tabular}[c]{@{}l@{}} SVM.\end{tabular} &
    \begin{tabular}[c]{@{}l@{}} 99\% \end{tabular}&
     \begin{tabular}[c]{@{}l@{}} 98\% \end{tabular}&
      \begin{tabular}[c]{@{}l@{}} 100\% \end{tabular}&
       \begin{tabular}[c]{@{}l@{}} 99\% \end{tabular}\\
     \begin{tabular}[c]{@{}l@{}}\end{tabular} &
  \begin{tabular}[c]{@{}l@{}} \end{tabular} &
  \begin{tabular}[c]{@{}l@{}}\end{tabular} &
  \begin{tabular}[c]{@{}l@{}} \end{tabular} &
   \begin{tabular}[c]{@{}l@{}} GNB.\end{tabular} &
    \begin{tabular}[c]{@{}l@{}} 84\% \end{tabular}&
    \begin{tabular}[c]{@{}l@{}} 100\% \end{tabular}&
    \begin{tabular}[c]{@{}l@{}} 68\% \end{tabular}&
    \begin{tabular}[c]{@{}l@{}} 80\% \end{tabular}\\
    
 \begin{tabular}[c]{@{}l@{}}\end{tabular} &
  \begin{tabular}[c]{@{}l@{}} \end{tabular} &
  \begin{tabular}[c]{@{}l@{}}\end{tabular} &
  \begin{tabular}[c]{@{}l@{}} \end{tabular} &
   \begin{tabular}[c]{@{}l@{}} MLP.\end{tabular} &
    \begin{tabular}[c]{@{}l@{}} 93\% \end{tabular}&
    \begin{tabular}[c]{@{}l@{}} 98\% \end{tabular}&
    \begin{tabular}[c]{@{}l@{}} 88\% \end{tabular}&
    \begin{tabular}[c]{@{}l@{}} 91\% \end{tabular}\\ \hline
   
     \begin{tabular}[c]{@{}l@{}}\end{tabular} &
  \begin{tabular}[c]{@{}l@{}}\end{tabular} &
  \begin{tabular}[c]{@{}l@{}}\end{tabular} &
  \begin{tabular}[c]{@{}l@{}} \end{tabular} &
   \begin{tabular}[c]{@{}l@{}} DT.\end{tabular} &
    \begin{tabular}[c]{@{}l@{}} 97\% \end{tabular}&
    \begin{tabular}[c]{@{}l@{}} 98\% \end{tabular}&
    \begin{tabular}[c]{@{}l@{}} 96\% \end{tabular}&
    \begin{tabular}[c]{@{}l@{}} 97\% \end{tabular}\\
    \begin{tabular}[c]{@{}l@{}}\end{tabular} &
  \begin{tabular}[c]{@{}l@{}} \end{tabular} &
  \begin{tabular}[c]{@{}l@{}}\end{tabular} &
  \begin{tabular}[c]{@{}l@{}} \end{tabular} &
   \begin{tabular}[c]{@{}l@{}} KNN.\end{tabular} &
    \begin{tabular}[c]{@{}l@{}} 99\% \end{tabular}&
    \begin{tabular}[c]{@{}l@{}} 98\% \end{tabular}&
    \begin{tabular}[c]{@{}l@{}} 100\% \end{tabular}&
    \begin{tabular}[c]{@{}l@{}} 99\% \end{tabular}\\
     \begin{tabular}[c]{@{}l@{}}GSE77314 \end{tabular} &
  \begin{tabular}[c]{@{}l@{}} Binary class  \end{tabular} &
  \begin{tabular}[c]{@{}l@{}}FGS \end{tabular} &
  \begin{tabular}[c]{@{}l@{}}12 \end{tabular} &
   \begin{tabular}[c]{@{}l@{}} SVM.\end{tabular} &
    \begin{tabular}[c]{@{}l@{}} 99\% \end{tabular}&
    \begin{tabular}[c]{@{}l@{}} 98\% \end{tabular}&
    \begin{tabular}[c]{@{}l@{}} 100\% \end{tabular}&
    \begin{tabular}[c]{@{}l@{}} 99\% \end{tabular}\\
     \begin{tabular}[c]{@{}l@{}}\end{tabular} &
  \begin{tabular}[c]{@{}l@{}} \end{tabular} &
  \begin{tabular}[c]{@{}l@{}}\end{tabular} &
  \begin{tabular}[c]{@{}l@{}} \end{tabular} &
   \begin{tabular}[c]{@{}l@{}} GNB.\end{tabular} &
    \begin{tabular}[c]{@{}l@{}} 97\% \end{tabular}&
    \begin{tabular}[c]{@{}l@{}} 98\% \end{tabular}&
    \begin{tabular}[c]{@{}l@{}} 96\% \end{tabular}&
    \begin{tabular}[c]{@{}l@{}} 96.8\% \end{tabular}\\
 \begin{tabular}[c]{@{}l@{}}\end{tabular} &
  \begin{tabular}[c]{@{}l@{}} \end{tabular} &
  \begin{tabular}[c]{@{}l@{}}\end{tabular} &
  \begin{tabular}[c]{@{}l@{}} \end{tabular} &
   \begin{tabular}[c]{@{}l@{}} MLP.\end{tabular} &
    \begin{tabular}[c]{@{}l@{}} 99\% \end{tabular}&
    \begin{tabular}[c]{@{}l@{}}98\% \end{tabular}&
    \begin{tabular}[c]{@{}l@{}} 100\% \end{tabular}&
    \begin{tabular}[c]{@{}l@{}}99\% \end{tabular}\\ \hline
    
    \begin{tabular}[c]{@{}l@{}}\end{tabular} &
  \begin{tabular}[c]{@{}l@{}}\end{tabular} &
  \begin{tabular}[c]{@{}l@{}}\end{tabular} &
  \begin{tabular}[c]{@{}l@{}} \end{tabular} &
   \begin{tabular}[c]{@{}l@{}} DT.\end{tabular} &
    \begin{tabular}[c]{@{}l@{}} 85.8\% \end{tabular}&
    \begin{tabular}[c]{@{}l@{}}83\% \end{tabular} &
    \begin{tabular}[c]{@{}l@{}} 82.6\% \end{tabular} &
    \begin{tabular}[c]{@{}l@{}} 81.5\%\end{tabular} \\
    \begin{tabular}[c]{@{}l@{}}\end{tabular} &
  \begin{tabular}[c]{@{}l@{}} \end{tabular} &
  \begin{tabular}[c]{@{}l@{}}\end{tabular} &
  \begin{tabular}[c]{@{}l@{}} \end{tabular} &
   \begin{tabular}[c]{@{}l@{}} KNN.\end{tabular} &
    \begin{tabular}[c]{@{}l@{}} 85\%
    \end{tabular}&
  \begin{tabular}[c]{@{}l@{}} 87.9\%
    \end{tabular}&
    \begin{tabular}[c]{@{}l@{}} 87.7\%
    \end{tabular}&
    \begin{tabular}[c]{@{}l@{}} 87\%
    \end{tabular}\\
     \begin{tabular}[c]{@{}l@{}}GSE45827 \end{tabular} &
  \begin{tabular}[c]{@{}l@{}} Multiclass \end{tabular} &
  \begin{tabular}[c]{@{}l@{}}No\end{tabular} &
  \begin{tabular}[c]{@{}l@{}}29873 \end{tabular} &
   \begin{tabular}[c]{@{}l@{}} SVM.\end{tabular} &
    \begin{tabular}[c]{@{}l@{}} 94.8\% \end{tabular}&
     \begin{tabular}[c]{@{}l@{}} 96\% \end{tabular}&
      \begin{tabular}[c]{@{}l@{}} 95.8\% \end{tabular}&
       \begin{tabular}[c]{@{}l@{}} 95.8\% \end{tabular}\\
     \begin{tabular}[c]{@{}l@{}}\end{tabular} &
  \begin{tabular}[c]{@{}l@{}} \end{tabular} &
  \begin{tabular}[c]{@{}l@{}}\end{tabular} &
  \begin{tabular}[c]{@{}l@{}} \end{tabular} &
   \begin{tabular}[c]{@{}l@{}} GNB.\end{tabular} &
    \begin{tabular}[c]{@{}l@{}} 89\% \end{tabular}&
    \begin{tabular}[c]{@{}l@{}} 92.7\% \end{tabular}&
    \begin{tabular}[c]{@{}l@{}} 88.8\% \end{tabular}&
    \begin{tabular}[c]{@{}l@{}} 89\% \end{tabular}\\
    
 \begin{tabular}[c]{@{}l@{}}\end{tabular} &
  \begin{tabular}[c]{@{}l@{}} \end{tabular} &
  \begin{tabular}[c]{@{}l@{}}\end{tabular} &
  \begin{tabular}[c]{@{}l@{}} \end{tabular} &
   \begin{tabular}[c]{@{}l@{}} MLP.\end{tabular} &
    \begin{tabular}[c]{@{}l@{}} 20.6\% \end{tabular}&
    \begin{tabular}[c]{@{}l@{}} 6\% \end{tabular}&
    \begin{tabular}[c]{@{}l@{}} 17\% \end{tabular}&
    \begin{tabular}[c]{@{}l@{}} 7\% \end{tabular} \\ \hline
     \begin{tabular}[c]{@{}l@{}}\end{tabular} &
  \begin{tabular}[c]{@{}l@{}}\end{tabular} &
  \begin{tabular}[c]{@{}l@{}}\end{tabular} &
  \begin{tabular}[c]{@{}l@{}} \end{tabular} &
   \begin{tabular}[c]{@{}l@{}} DT \end{tabular} &
    \begin{tabular}[c]{@{}l@{}} 89.6\% \end{tabular}&
    \begin{tabular}[c]{@{}l@{}} 90.9\% \end{tabular}&
    \begin{tabular}[c]{@{}l@{}} 89.6\% \end{tabular}&
    \begin{tabular}[c]{@{}l@{}} 88.8\% \end{tabular}\\
    \begin{tabular}[c]{@{}l@{}}\end{tabular} &
  \begin{tabular}[c]{@{}l@{}} \end{tabular} &
  \begin{tabular}[c]{@{}l@{}}\end{tabular} &
  \begin{tabular}[c]{@{}l@{}} \end{tabular} &
   \begin{tabular}[c]{@{}l@{}} KNN\end{tabular} &
    \begin{tabular}[c]{@{}l@{}} 95.48\%
    \end{tabular}&
      \begin{tabular}[c]{@{}l@{}} 96.5\%
    \end{tabular}&
      \begin{tabular}[c]{@{}l@{}} 96\%
    \end{tabular}&
      \begin{tabular}[c]{@{}l@{}} 96\%
    \end{tabular}\\
    
     \begin{tabular}[c]{@{}l@{}}GSE45827 \end{tabular} &
  \begin{tabular}[c]{@{}l@{}} Multiclass \end{tabular} &
  \begin{tabular}[c]{@{}l@{}}FGS \end{tabular} &
  \begin{tabular}[c]{@{}l@{}}68 \end{tabular} &
   \begin{tabular}[c]{@{}l@{}} SVM.\end{tabular} &
    \begin{tabular}[c]{@{}l@{}} 98.7\% \end{tabular}&
    \begin{tabular}[c]{@{}l@{}} 99\% \end{tabular}&
    \begin{tabular}[c]{@{}l@{}} 98.8\% \end{tabular}&
    \begin{tabular}[c]{@{}l@{}} 98.9\% \end{tabular}\\
     \begin{tabular}[c]{@{}l@{}}\end{tabular} &
  \begin{tabular}[c]{@{}l@{}} \end{tabular} &
  \begin{tabular}[c]{@{}l@{}}\end{tabular} &
  \begin{tabular}[c]{@{}l@{}} \end{tabular} &
   \begin{tabular}[c]{@{}l@{}} GNB.\end{tabular} &
    \begin{tabular}[c]{@{}l@{}} 91.6\% \end{tabular}&
     \begin{tabular}[c]{@{}l@{}} 94.5\% \end{tabular}&
      \begin{tabular}[c]{@{}l@{}} 92\% \end{tabular}&
       \begin{tabular}[c]{@{}l@{}} 92.8\% \end{tabular}\\
 \begin{tabular}[c]{@{}l@{}}\end{tabular} &
  \begin{tabular}[c]{@{}l@{}} \end{tabular} &
  \begin{tabular}[c]{@{}l@{}}\end{tabular} &
  \begin{tabular}[c]{@{}l@{}} \end{tabular} &
   \begin{tabular}[c]{@{}l@{}} MLP.\end{tabular} &
    \begin{tabular}[c]{@{}l@{}} 98.7\% \end{tabular}&
    \begin{tabular}[c]{@{}l@{}} 99.3\% \end{tabular}&
    \begin{tabular}[c]{@{}l@{}} 98.8\% \end{tabular}&
    \begin{tabular}[c]{@{}l@{}} 98.9\% \end{tabular}\\ \hline
\end{tabular}
 \end{table*}\\
 
 \subsection{Results discussion}
\begin{figure}[!tbp]
  \centering
  \begin{minipage}[b]{0.45\textwidth}
    \includegraphics[width=\textwidth]{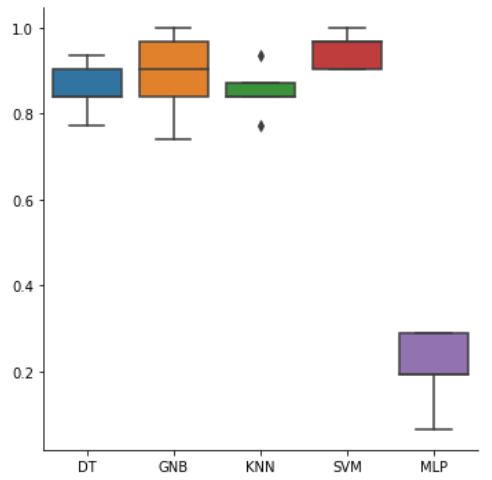}
    \caption{Accuracy scores for breast cancer (GSE45827) before employing FGS}
    \label{fig6}
  \end{minipage}
  \hfill
  \begin{minipage}[b]{0.45\textwidth}
    \includegraphics[width=\textwidth]{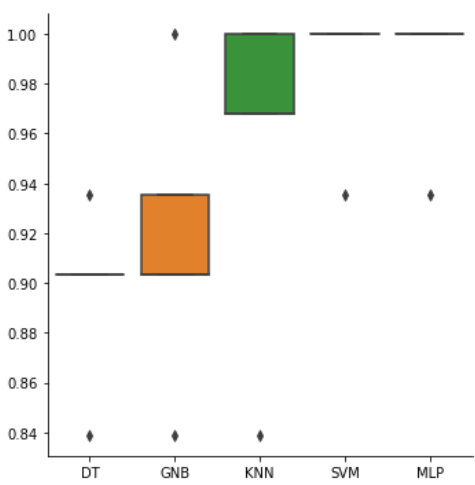}
    \caption{Accuracy scores for breast cancer (GSE45827) when employing FGS}
    \label{fig7}
  \end{minipage}
\end{figure}

\begin{figure}[!tbp]
  \centering
  \begin{minipage}[b]{0.45\textwidth}
    \includegraphics[width=\textwidth]{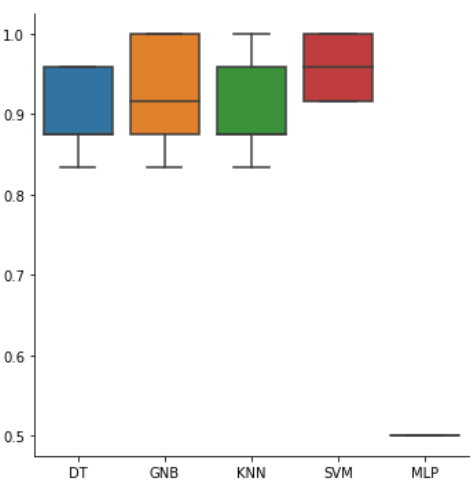}
    \caption{Accuracy scores for lung cancer (GSE19804)
without applying FGS}
\label{fig10}
  \end{minipage}
  \hfill
  \begin{minipage}[b]{0.45\textwidth}
    \includegraphics[width=\textwidth]{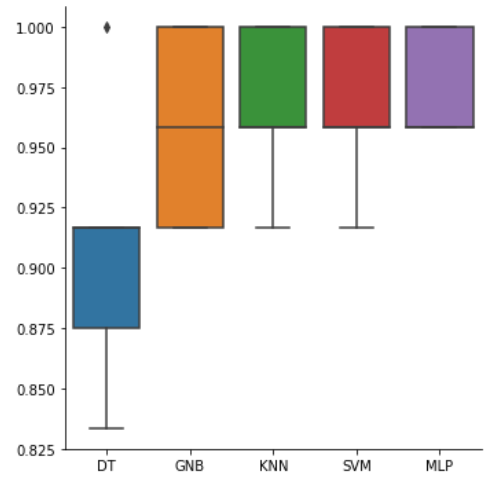}
    \caption{Accuracy scores for lung cancer (GSE19804) when FGS method applied}
    \label{fig11}
  \end{minipage}
\end{figure}

\begin{figure}[!tbp]
  \centering
  \begin{minipage}[b]{0.45\textwidth}
    \includegraphics[width=\textwidth]{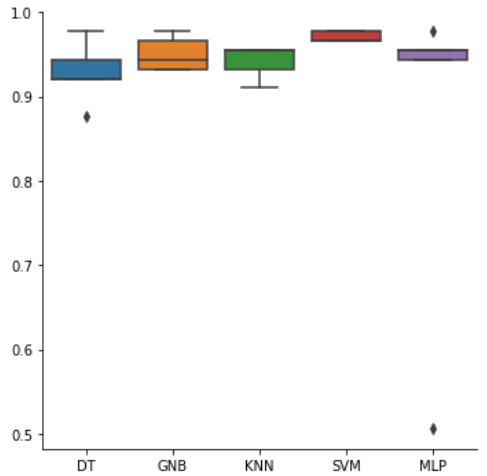}
    \caption{Accuracy scores for liver cancer (GSE14520 )
without applying FGS}
\label{fig12}
  \end{minipage}
  \hfill
  \begin{minipage}[b]{0.45\textwidth}
    \includegraphics[width=\textwidth]{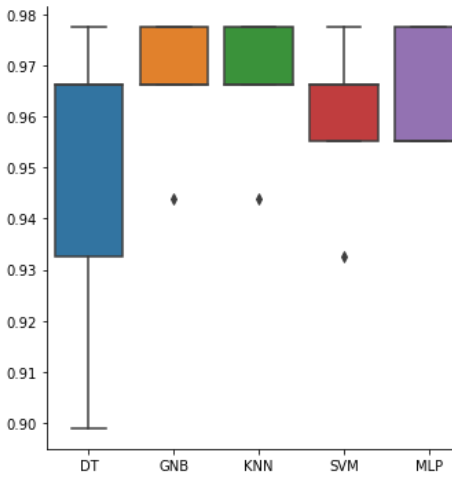}
    \caption{Accuracy scores for liver cancer (GSE14520)
when FGS method applied}
\label{fig13}
  \end{minipage}
\end{figure}

\begin{figure}[!tbp]
  \centering
  \begin{minipage}[b]{0.45\textwidth}
    \includegraphics[width=\textwidth]{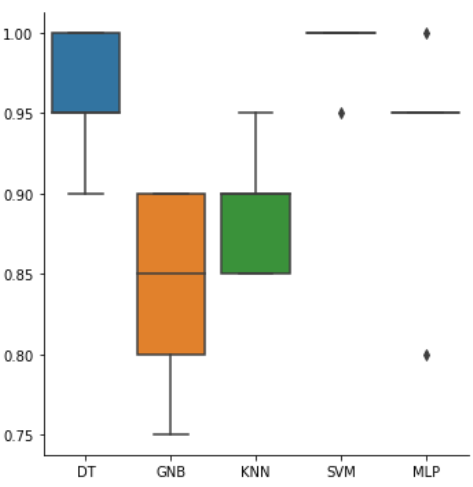}
    \caption{Accuracy score for liver cancer (GSE77314) in 5 kfolds without using FGS}
    \label{fig20}
  \end{minipage}
  \hfill
  \begin{minipage}[b]{0.45\textwidth}
    \includegraphics[width=\textwidth]{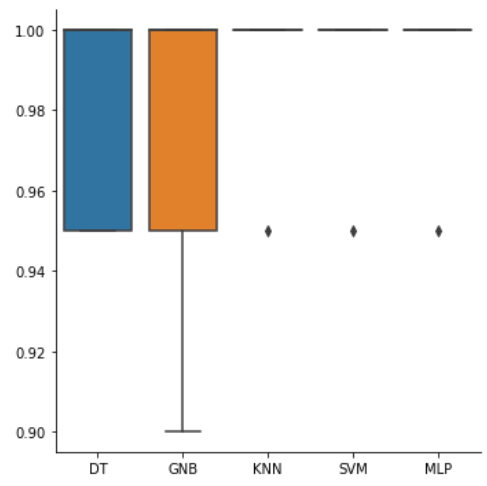}
    \caption{Accuracy score for liver cancer (GSE77314) in 5 kfolds when FGS used}
    \label{fig30}
  \end{minipage}
\end{figure}

\begin{figure}[!tbp]
  \centering
  \begin{minipage}[b]{0.45\textwidth}
    \includegraphics[width=\textwidth]{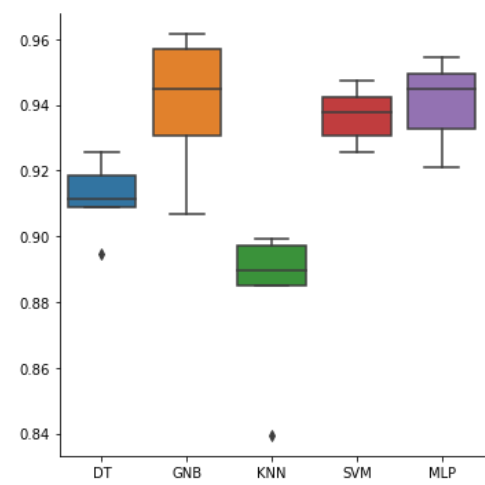}
    \caption{Accuracy scores in 5 kfolds for the (TCGA) datasets without applying FGS}
    \label{fig14}
  \end{minipage}
  \hfill
  \begin{minipage}[b]{0.45\textwidth}
    \includegraphics[width=\textwidth]{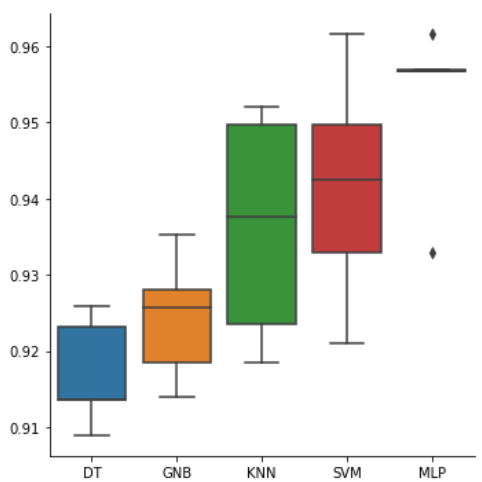}
    \caption{Accuracy scores in 5 kfolds for the (TCGA) datasets when  FGS employed}
    \label{fig15}
  \end{minipage}
\end{figure}

\begin{figure}[!tbp]
  \centering
  \begin{minipage}[b]{0.45\textwidth}
    \includegraphics[width=\textwidth]{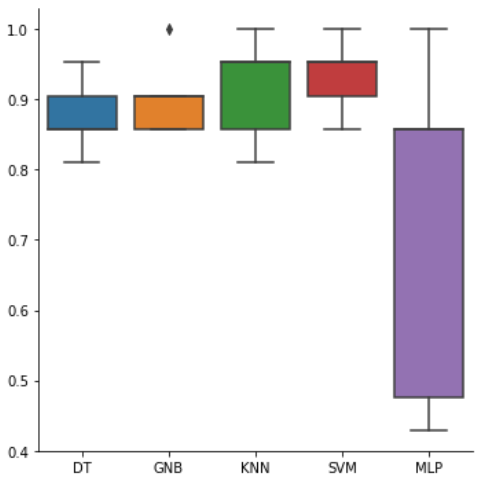}
    \caption{Accuracy score in 5 kfolds for thyroid cancer (GSE33630) by omitting FGS}
    \label{fig17}
  \end{minipage}
  \hfill
  \begin{minipage}[b]{0.45\textwidth}
    \includegraphics[width=\textwidth]{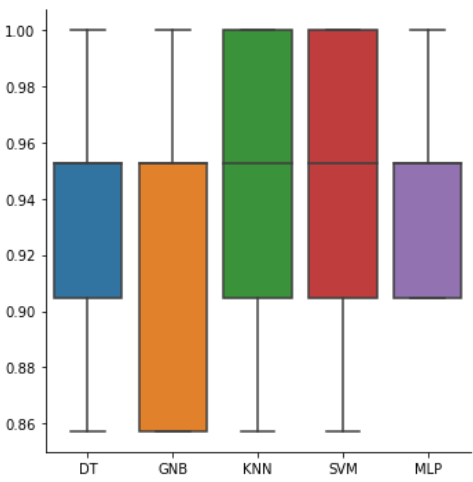}
    \caption{Accuracy score in 5 kfolds for thyroid cancer (GSE33630) when FGS used.}
    \label{fig18}
  \end{minipage}
\end{figure}

To show the differences between the results obtained by omitting and employing the FGS technique with the five different classifier techniques, the accuracy scores in 5 kfolds have been displayed on a bar chart. The two bar graphs (\ref{fig6} and \ref{fig7})  demonstrate the five-fold difference in accuracy ratings between utilizing and ignoring FGS. The two bar graphs demonstrate how the usage of FGS enhanced classifier model performance, notably with the MLP classifier. The FGS method was also utilized to reduce the number of selected genes from 29873 to 68 genes. These results suggest that the development of the FGS technique contributed to an improvement in accuracy, a reduction in the training time for models, and the provision of early cancer detection by the choice of instructive genes. Classifier models are also less complicated.\\
As shown in the two bar charts (\ref{fig10} and \ref{fig11}), a fuzzy gene selection strategy significantly improved the performance of the five classifier approaches for classifying lung cancer. In comparison to other classifier models, the findings demonstrate that the MLP model offers predictions that are closer to the ideal observed value. MLP earned an average accuracy score of 97.5 in 5 kfolds. Other classifiers, however, achieved average scores of 96.6, 96.6, 95.8, and 92.5 in 5 kfolds for SVM, KNN, GNB, and DT, respectively. Additionally, only 36 genes out of 45782 genes were employed for training the classifier models, a considerable decrease in the number of genes used.
\\
Although there is a slight improvement in the accuracy of most of the classifiers used in this study to classify liver cancer datasets(GSE14520). However, there is a significant enhancement in the MLP classifier when using the FGS method, as it improved from 86.6 to 96 as an average accuracy score in 5 kfolds. More importantly, the FGS method reduced the number of genes used to train models to 23 only out of 13425. The two bar charts (\ref{fig12} and \ref{fig13}) explain the comparison accuracy scores with 5 kfolds for the five models when FGS employed and omitted.\\
\\
Most classifier models used reached close to 100 where the average accuracy score in 5 kfolds is 99\% for the SVM, KNN, and MLP while 97\% for GNB and DT when fuzzy gene selection techniques are applied to the liver cancer dataset (GSE77314). These remarkable enhancements in accuracy score are shown in (\ref{fig20} and \ref{fig30}). Moreover, the FGS method decreased the number of genes from 29087 to only 12 genes that were used as identifiers for training the proposed model and compared models. That leads to an increase in the model efficiency and mitigates the time taken through algorithm training and provides early cancer detection.\\
\\
There was not a significant improvement in (TCGA) datasets because the number of genes used was not large (971), so its use did not achieve a high level of accuracy improvement. However, it improved the performance of the model by reducing the number of selected genes that were used as identifiers to train the technique. As a result, the FGS method decreased the number of genes from 971 to 25 genes only. In addition, a slight improvement in the accuracy as well as the precision, we conclude that employing FGS in the worst cases will give better accuracy and fewer genes, and that performed less time for training the classifier models and provides early detection of cancer. The two bar charts (\ref{fig14} and \ref{fig15}) illustrate the difference between the accuracy scores in 5 kfolds when the classifier models were applied to the datasets with omitting FGS and the accuracy score in 5 kfolds when the classifier applied to the selected genes by FGS method.\\
\\
For the majority of applied classifier models, and specifically, MLP, where 72\% is the average accuracy score in 5 kfolds when omitting FGS, while 93\% when FGS is employed, good enhancement is obtained when the fuzzy gene selection method is applied to thyroid cancer (GSE33630) datasets. Additionally, the number of genes was reduced from 23516 to 76 genes, which reduced the complexity, interpretability, and training time for algorithms as well as enabled the early identification of cancer. The two bar graphs (\ref{fig17} and \ref{fig18}) show the differences in accuracy scores for five distinct classifier models when the FGS approach is used in comparison to when it is not used.
\\
Briefly, multilayer perceptron achieved the highest average accuracy across the six datasets when fuzzy gene selection was applied which was 96.5\%. It also, MLP has accomplished the highest improvement rate for the average accuracy when the proposed fuzzy gene selection which was 27.3\% . It can be concluded that the highest improvement impact of fuzzy gene selection was when a MLP classifier was employed and the accuracy improved from 69.2\% before FGS was applied while 96.5\% when FGS was applied.\\
Based on the results that were explained previously, a full automated deep neural network was proposed to analyze gene expression data as described in (Figure \ref{fig:Fig9}).  The proposed model attempted to achieve three main goals as follows: The first goal, reducing the number of genes that would be used as identifiers for training a classifier method in resulting that leads to reduce the time consuming of training a model. Indeed, the proposed model succeeded remarkably in reducing the number of genes as indicated in (Table \ref{tab3} and Table \ref{tab4}). The second goal, enhancing the performance of the accuracy and other evaluation measurement parameters and the aim was also accomplished where the average accuracy was 96.5\%. The third goal, selecting candidate genes as putative targets for biologists to further investigate to determine whether these genes simply useful for classification or are implicated in the pathogenesis of these diseases.
\section{Conclusion}\label{sec13}

In order to improve the machine learning performance for cancer classification, this research introduces a novel fuzzy gene selection approach for lowering the dimensionality (reducing the number of features) of gene expression data. It also decreases the amount of time needed for algorithm training. Using the commonly used measurement techniques ( Microarray and RNA-seq) for estimating gene expression data, the proposed model was trained and evaluated on six datasets obtained from TCGA and GEO. Three primary objectives were accomplished by this work: to boost the effectiveness of classifier techniques, help speed up the training process and cut down on the number of chosen genes that are utilized as identifiers for the classifier training model. The findings demonstrate that the suggested model (FGS-MLP) has the best accuracy in the majority of the datasets studied, with accuracy levels ranging from 93\% at the lowest end to 99\% at the top. 

The average accuracy rating across six datasets is 96.5\%. As a result, the proposed model shows both the capacity to properly classify cancer and time savings during the training phase. By more carefully choosing characteristics (genes) from different cancer kinds, biologists  can also benefit from the selected genes in their study and early cancer detection. Furthermore, FGS may also assist in reducing the complexity of a classifier method and avoiding or at least mitigating the overfitting issue that typically arises when high dimensionality datasets are used.

Regardless of the contributions and promising findings
of this research, it has some limitations. First, a limited
number of datasets used that can more datasets used for
different cancer types especially RNA-seq data. Additionally,
no single classical ML classifier can continuously achieve the
best accuracy in all given datasets. Due to these limitations,
future work will make an effort to use more datasets for
different cancer types and propose a new classifier that can
accurately and continuously classify gene expression data.
\section*{Declarations}

\begin{itemize}
\item Funding
This research was partly funded by the Ministry of Higher Education and Scientific Research in the Republic of Iraq, according to 
scholarship number (22223) on (06/09/2017) to sponsor the first author to pursue his PhD research.
\item Conflict of interest/Competing interests (check journal-specific guidelines for which heading to use).
\\
Not applicable
\item Ethics approval 
\\
Not applicable
\item Consent to participate
\item Consent for publication
\item Availability of data and materials.
 The original datasets that were employed for cancer classification are freely available 
at:\url{https://github.com/mahmoodjasim/OrginalDataset}. While the final datasets that have been used after applying Fuzzy gene selection method are freely available at:\url{https://github.com/mahmoodjasim/Datasets-of-selected-genes}
\item Code availability.\\
 The codes used in this article are freely available at:\url{ https://github.com/mahmoodjasim/Fuzzy-Gene-Selection-Code}
\item Authors' contributions
\end{itemize}

\end{document}